\documentclass[prd,preprint,superscriptaddress,amsmath,amssymb,nofootinbib]{revtex4}
\pdfoutput=1

\usepackage{graphicx,color,slashed}
\usepackage{subfigure}
\def\be{\begin{eqnarray}}
\def\ed{\end{eqnarray}}

\def\be{\begin{eqnarray}}
\def\ed{\end{eqnarray}}
\def\beq{\begin{equation}}
\def\eeq{\end{equation}}
\def\bea{\begin{eqnarray}}
\def\eea{\end{eqnarray}}

\begin{document}


\title{\bf \Large CP asymmetries of ${\overline B}\to X_s/X_d\gamma$ in models with\\ 
three Higgs doublets}

\author{A.G. Akeroyd}
\email{a.g.akeroyd@soton.ac.uk}
\affiliation{School of Physics and Astronomy, University of Southampton,
Highfield, Southampton SO17 1BJ, United Kingdom}

\author{Stefano Moretti}
\email{S.Moretti@soton.ac.uk}

\affiliation{School of Physics and Astronomy, University of Southampton,
Highfield, Southampton SO17 1BJ, United Kingdom}

\author{Tetsuo Shindou}
\email{shindou@cc.kogakuin.ac.uk}

\affiliation{Division of Liberal-Arts, Kogakuin University,
2665-1 Nakano-machi, Hachioji, Tokyo, 192-0015, Japan}

\author{Muyuan Song}
\email{ms32g13@soton.ac.uk}

\affiliation{School of Physics and Astronomy, University of Southampton,
Highfield, Southampton SO17 1BJ, United Kingdom}

\date{\today}

\begin{abstract}
\noindent
Direct CP asymmetries (${\cal A}_{CP}$) in the inclusive decays of ${\overline B}\to X_s\gamma$ and ${\overline B}\to X_{s+d}\gamma$ of the order of $1\%$ will be probed at the BELLE II experiment.  In this work, three such asymmetries are studied in the context of a three-Higgs-doublet model (3HDM), and it is shown that all three ${\cal A}_{CP}$ can be as large as the current experimental limits. 
Of particular interest is ${\cal A}_{CP}$ for ${\overline B}\to X_{s+d}\gamma$, which is predicted to be effectively zero in the Standard Model (SM). 
A measurement of $2.5\%$ or more for this observable with the full BELLE II data 
would give $5\sigma$ evidence for physics beyond the SM. We display parameter space in the 3HDM for which such a clear signal is possible.

\end{abstract}

\maketitle

\section{Introduction}
\noindent
A new particle  with a mass of around 125 GeV was discovered by the ATLAS and CMS collaborations of the Large Hadron Collider (LHC) 
\cite{Aad:2012tfa,Chatrchyan:2012xdj}.
At present, the measurements of its properties are in very good agreement 
(within experimental error) with those of the Higgs boson of the 
Standard Model (SM), and hence the simplest interpretation is that the 125 GeV scalar boson is the 
(lone) Higgs boson of the SM ($h$).
However, an alternative possibility is that 
it is the first scalar to be discovered from a non-minimal Higgs 
sector, which contains additional scalar isospin doublets 
or higher representations (e.g. scalar isospin triplets). 
This scenario will be probed by more precise future measurements of its branching ratios (BRs), which might
eventually show deviations from 
those of the SM Higgs boson. There would also be the possibility of 
discovering additional electrically neutral scalars ($H$ or $A$) and/or electrically charged scalars ($H^\pm$), and such searches form an
active part of the LHC experimental programme. 
In the context of a two-Higgs-doublet model (2HDM) 
the lack of direct observation of 
an $H^\pm$ at the LHC together with precise measurements of SM processes exclude parameter space of $\tan\beta$ (which is
present in the Yukawa couplings)
and $m_{H^\pm}$ (mass of the $H^\pm$), where $\tan\beta=v_2/v_1$, 
and $v_1$ and $v_2$ are the vacuum expectation values (VEVs) of the two 
Higgs doublets, respectively (for reviews  see e.g. \cite{Branco:2011iw},\cite{Akeroyd:2016ymd}). 

In a three-Higgs-doublet model (3HDM) the Yukawa couplings of the two 
charged scalars depend on the four free parameters ($\tan\beta$, $\tan\gamma$,
$\theta$, and $\delta$) of a unitary matrix 
that rotates the charged scalar fields in the weak eigenbasis to the
physical charged scalar fields \cite{Weinberg:1976hu}.  The phenomenology of the lightest $H^\pm$ in a 3HDM
\cite{Grossman,Akeroyd:1994ga,Akeroyd2} can be different to that of $H^\pm$ in a 2HDM due to the larger
number of parameters that determine its fermionic couplings.

The decay $b\to s\gamma$, whose BR has been measured to be in good agreement with that of the SM, provides strong constraints on the parameter space of charged scalars in 2HDMs and 3HDMs. 
In the well-studied 2HDM Type II the bound $m_{H^\pm}> 480$ GeV \cite{Misiak:2015xwa} can be obtained and is valid for all $\tan\beta$. More precise measurements of BR$(b\to s\gamma)$
at the ongoing BELLE II experiment will sharpen these constraints, but it is very unlikely that measurements of BR$(b\to s\gamma)$ alone could provide evidence for the existence of an $H^\pm$. However, the direct CP asymmetry in $b\to s\gamma$ 
will be probed at the 1\% level,
and can be enhanced significantly above the SM prediction by additional complex phases that are present in models of physics beyond the SM \cite{Kagan:1998bh}.
In the context of 3HDMs we study the magnitude of three different direct CP asymmetries that involve $b\to s\gamma$, including the contribution of both charged scalars for the
first time. We display parameter space in 3HDMs that would give a clear signal for these three observables at the BELLE II experiment.

This work is organised as follows. 
In section II the measurements of $b\to s\gamma$ are summarised and the CP asymmetries in this
decay are described. In section III the contribution of the charged scalars in a 3HDM to the partial decay width of 
$b\to s(d)\gamma$ is presented. Section IV contains our results, and conclusions are given in 
section V.

\section{Direct CP asymmetries in ${\overline B}\to X_s \gamma$ and ${\overline B}\to X_{s+d} \gamma$}
\noindent
In this section the experimental measurements of  the inclusive decays ${\overline B}\to X_s \gamma$ and ${\overline B}\to X_{s+d} \gamma$ (charged conjugated processes are implied) are described, followed by a 
discussion of direct CP asymmetries in these decays. The symbol $B$ signifies $B^+$ or $B^0$ (which contain anti-$b$ quarks), while $\overline B$ signifies 
$B^-$ or $\overline{B^0}$ (which contain $b$ quarks). The symbol
$X_{s}$ denotes any hadronic final state that originates from a strange quark hadronising (e.g. states with at least one kaon), $X_{d}$ means any hadronic final state 
that originates from a down quark hadronising (e.g. states with at least one pion), and $X_{s+d}$ denotes any hadronic final state that is $X_{s}$ or $X_{d}$. 

\subsection{Experimental measurements of ${\overline B}\to X_s \gamma$ and ${\overline B}\to X_{s+d} \gamma$}
\noindent
There are two ways to measure the BR of the inclusive decays ${\overline B}\to X_{s/d} \gamma$:\\
i) The fully inclusive method;\\
ii) The sum-of-exclusives method (also known as "semi-inclusive"). 

In the fully inclusive approach only a photon from the signal $\overline B$ (or $B$) meson in the $B\overline B$ event, which decays via $b\to s/d\gamma$, is selected. Consequently, 
this method cannot distinguish between hadronic states $X_s$ and $X_d$, and what is measured is actually the sum of ${\overline B}\to X_s\gamma$ and 
${\overline B}\to X_d \gamma$. From the other $\overline B$ (or $B$) meson ("tag $B$ meson") either a 
lepton ($e$ or $\mu$) can be selected or full reconstruction (either hadronic or semi-leptonic) can be carried out. The former method has a higher signal efficiency, but the latter method has greater background suppression. 
Measurements of  ${\overline B}\to X_{s+d} \gamma$ using the fully inclusive method with leptonic tagging have been carried out by 
the CLEO collaboration  \cite{Chen:2001fja}, 
the BABAR collaboration \cite{Lees:2012ufa} and the BELLE collaboration \cite{Belle:2016ufb}.
A measurement of ${\overline B}\to X_{s+d} \gamma$ using the fully inclusive method with full (hadronic) reconstruction  of the tag ${\overline B}$ meson has so far only been
carried out by the BABAR collaboration \cite{Aubert:2007my}. At the current integrated luminosities (0.5 to 1 ab$^{-1}$) the errors associated with measurements
that involve full reconstruction are significantly larger than the errors from 
measurements with leptonic tagging. However,  with the larger integrated luminosity at BELLE II (50 ab$^{-1}$)
it is expected that both approaches will provide roughly similar errors. To obtain a measurement of ${\overline B}\to X_s\gamma$ alone, the contribution of ${\overline B}\to X_d \gamma$
(which is smaller by roughly $|V_{td}/V_{ts}|^2 \approx 1/20$ in the SM, which has also been confirmed experimentally) is subtracted.

In the sum-of-exclusives approach the selection criteria are sensitive to as many exclusive decays as possible in the hadronic final states $X_s$ and $X_d$ 
of the signal ${\overline B}$, as well as requiring a photon from $b\to s/d \gamma$.  
In contrast to the fully inclusive approach, no selection is made on the other $B$ meson in the $B\overline B$ event.
The sum-of-exclusives method is sensitive to whether the decay $b\to s\gamma$ or $b\to d \gamma$ occurred and so this approach measures
${\overline B}\to X_s \gamma$ or ${\overline B}\to X_{d} \gamma$. It has different systematic uncertainties to that of the fully inclusive approach.
Measurements of ${\overline B}\to X_s \gamma$ have been carried out by the 
BABAR collaboration \cite{Lees:2012wg} and the BELLE collaboration \cite{Saito:2014das}.
Currently, 38 exclusive decays in ${\overline B}\to X_s \gamma$ (about 70\% of the total BR) and 7 exclusive decays
in ${\overline B}\to X_d \gamma$ \cite{delAmoSanchez:2010ae} are included.
At current integrated luminosities the error in the measurement of ${\overline B}\to X_s\gamma$ is about twice that of the fully inclusive approach, whereas at BELLE II
integrated luminosities the latter is still expected to give the more precise measurement. 

Measurements in both the above approaches are made with a lower cut-off on the photon energy $E_\gamma$ in the range 1.7 GeV to 2.0 GeV, and then an 
extrapolation is made to $E_{\gamma} > 1.6$ GeV using theoretical models.
The current world average for the above six measurements of ${\overline B}\to X_s \gamma$ is \cite{Amhis:2019ckw}:
\begin{equation}
\mathcal{B}^{\text{exp} }_{s\gamma}  =(3.32 \pm 0.15) \times 10^{-4}  \quad{\rm with}\quad E_{\gamma} > 1.6 \,\text{GeV}\,. \\ 
\label{bsy_exp}
\end{equation}
The error is currently $4.5\%$, and is expected to be reduced to around $2.6\%$ with the final integrated luminosity at the BELLE II experiment \cite{Kou:2018nap}.

The theoretical prediction including corrections to order $\alpha^2_s$ (i.e. Next-to-Next-to leading order, NNLO) is \cite{Misiak:2020vlo}:
\begin{equation}
\mathcal{B}^{SM}_{s\gamma} =(3.40\pm 0.17) \times 10^{-4} \quad{\rm with}\quad E_{\gamma} > 1.6\, \text{GeV} \,.\\ \nonumber
\end{equation}
There is excellent agreement between the world average and the NNLO prediction in the SM. Consequently, 
$\mathcal{B}^{\text{exp} }_{s\gamma} $ allows stringent lower limits to be derived on the mass of new particles, most notably 
the mass of the charged scalar ($m_{H^\pm}> 480$ GeV  \cite{Misiak:2015xwa}, as mentioned earlier)
in the 2HDM (Type II).

\subsection{Direct CP asymmetries of $\overline B\to X_s \gamma$ and $\overline B\to X_{s+d} \gamma$}
\noindent
Although it is clear that measurements of BR$({\overline B}\to X_s\gamma)$ alone will not provide evidence for new physics with BELLE II data,
the direct CP asymmetry in this decay might \cite{Kagan:1998bh}. 
Direct CP asymmetries in ${\overline B}\to X_s \gamma$ and ${\overline B}\to X_{d} \gamma$ are defined as follows:
\begin{equation}
\mathcal{A}_{X_{s(d)} \gamma}=\frac{\Gamma(\overline B\to X_{s(d)}\gamma)-\Gamma(B\to X_{{\overline s}({\overline d})}\gamma)}{\Gamma(\overline B\to X_{s(d)}\gamma)+\Gamma(B\to X_{{\overline s}({\overline d})}\gamma)}\,.
\end{equation}
If $B$ is $B^+$ (and so $\overline B=B^-$) in the definition of $\mathcal{A}_{X_{s(d)} \gamma}$  then the CP asymmetry is for the charged $B$ mesons, is labelled as $\mathcal{A}^\pm_{X_{s}\gamma}$ or $\mathcal{A}^\pm_{X_{d}\gamma}$,
and can be individually probed in a search that reconstructs $X_s$ or $X_d$  (sum-of-exclusives method). 
If $B$ is $B^0$ the CP asymmetry is for the neutral $B$ mesons, is labelled as $\mathcal{A}^0_{X_{s}\gamma}$ or $\mathcal{A}^0_{X_{d}\gamma}$, and can also be individually probed.
A general formula for the short-distance contribution (from "direct photons") to 
$\mathcal{A}_{X_{s(d)}\gamma}$  in terms of Wilson coefficients was derived in Ref.~\cite{Kagan:1998bh}. Prior to the publication of  Ref.~\cite{Kagan:1998bh} a few works 
\cite{Wolfenstein:1994jw,Asatrian:1996as,Borzumati:1998tg} had calculated $\mathcal{A}_{X_{s}\gamma}$ in the SM and in specific extensions of the it that include a charged Higgs boson. 
The formula for  $\mathcal{A}_{X_{s(d)}\gamma}$ in Ref.~\cite{Kagan:1998bh} was the first complete calculation of the asymmetry in terms of all the contributing Wilson coefficients, and
was extended twelve years later to include the long-distance contributions (from "resolved photons") in Ref.~\cite{Benzke:2010tq}.
In approximate form $\mathcal{A}_{X_{s(d)}\gamma}$  is as follows:
\bea \label{acpshortlong}
\mathcal{A}_{X_{s(d)}\gamma} &\approx & \pi \bigg \{  \bigg [ \bigg ( \frac{40}{81} - \frac{40}{9} \frac{\Lambda_c}{m_b} \bigg ) \frac{\alpha_s}{\pi} + \frac{\tilde{\Lambda}^{c}_{17}}{m_b} \bigg ] \text{Im} \frac{C_2}{C_{7\gamma}}   \nonumber \\&-& \bigg ( \frac{4\alpha_s}{9\pi} - 4\pi \alpha_s e_{\text{spec}} \frac{\tilde{\Lambda}_{78}}{m_b}  \bigg ) \text{Im} \frac{C_{8g}}{C_{7\gamma}} \nonumber \\ & - & \bigg ( \frac{\tilde{\Lambda}^{u}_{17} - \tilde{\Lambda}^{c}_{17} }{m_b} + \frac{40}{9} \frac{\Lambda_c}{m_b} \frac{\alpha_s}{\pi} \bigg) \text{Im} \bigg( \epsilon_{s(d)} \frac{C_2}{C_{7\gamma}} \bigg ) \bigg \}\,.
\eea
The above four asymmetries are obtained from eq.~(\ref{acpshortlong}) with the choices for $e_{\text{spec}}$ (the charge of the spectator quark) and $\epsilon_{s(d)}$ given in Tab.~\ref{asymmetries}.
\begin{table}[h]
\begin{center}
\begin{tabular}{|c||c|c|}
\hline
$\mathcal{A}_{X_{s(d)} \gamma}$ &  $e_{\text{spec}}$ & $\epsilon_{s(d)}$  \\ \hline
$\mathcal{A}^0_{X_s \gamma}$ &  $-\frac{1}{3}$ & $\epsilon_{s}$  \\ 
$\mathcal{A}^\pm_{X_s \gamma}$ & $\frac{2}{3}$ &$\epsilon_{s}$  \\
$\mathcal{A}^0_{X_d \gamma}$ & $-\frac{1}{3}$ & $\epsilon_{d}$   \\
$\mathcal{A}^\pm_{X_d \gamma}$ & $\frac{2}{3}$ & $\epsilon_{d}$  \\
\hline
\end{tabular}
\end{center}
\caption{The choices of $e_{\text{spec}}$ and $\epsilon_{s(d)}$ in the generic formula for $\mathcal{A}_{X_{s(d)} \gamma}$ that give rise to the four asymmetries.}
\label{asymmetries}
\end{table}
The parameters $\tilde{\Lambda}^{u}_{17} , \tilde{\Lambda}^{c}_{17} , \tilde{\Lambda}_{78}$ are hadronic parameters that determine the magnitude of the
long-distance contribution. Their allowed ranges were updated in Ref.~\cite{Gunawardana:2019gep} to be as follows:
\bea 
-660  \; \text{MeV} < \tilde{\Lambda}^{u}_{17} < + 660  \;\text{MeV}\,, \nonumber \\
-7  \; \text{MeV} < \tilde{\Lambda}^{c}_{17} <  + 10  \; \text{MeV} \,,\nonumber  \\
17  \;\text{MeV} <  \tilde{\Lambda}_{78} <190  \; \text{MeV}\,.
\label{hardonic_parameter}
\eea
The short-distance contributions to $\mathcal{A}_{X_{s(d)}\gamma}$ are the terms that are independent of $\Lambda_{ij}$, and $\mathcal{A}^0_{X_{s(d)}\gamma}=\mathcal{A}^\pm_{X_{s(d)}\gamma}$  if long-distance
terms are neglected.
Other parameters are as follows: $\Lambda_c = 0.38 \; \text{GeV}$,
$\epsilon_s = (V_{ub} V^*_{us})/ (V_{tb}V^*_{ts}) = \lambda^2 (i\bar{\eta} - \bar{\rho})/ [1- \lambda^2 (1 - \bar{\rho} + i \bar{\eta})]$ (in terms of Wolfenstein parameters), $\epsilon_d = (V_{ub} V^*_{ud})/ (V_{tb}V^*_{td}) = (\bar{\rho}-i\bar{\eta})/ (1- \bar{\rho} + i \bar{\eta})$. 
The $C_i$'s are Wilson coefficients 
of relevant operators that are listed in Ref.~\cite{Kagan:1998bh}. 
In the SM the Wilson coefficients are real and the only term in $\mathcal{A}_{X_{s(d)}\gamma}$ that is non-zero is the term with $\epsilon_{s(d)}$.
Due to $\epsilon_s$ being of order 
 $\lambda^2$ while $\epsilon_d$ is of order 1, for the imaginary parts one has ${\rm Im}(\epsilon_d)/{\rm Im}(\epsilon_s)\approx -22$.
For the short-distance contribution only (i.e. neglecting the term with $(\Lambda^{u}_{17}-\Lambda^{c}_{17})/{m_b}$ in eq.~(\ref{acpshortlong})) one has
$\mathcal{A}_{X_{s}\gamma}\approx 0.5\%$ and $\mathcal{A}_{X_{d}\gamma}\approx 10\%$. The small value of $\mathcal{A}_{X_{s}\gamma}$ in the SM
suggests that this observable could probe models of physics beyond the SM that contain Wilson coefficients with an imaginary part.

After the publication of Ref.~\cite{Kagan:1998bh},
several works calculated $\mathcal{A}_{X_{s}\gamma}$  (for the short-distance contribution only)
in the context of specific models of physics beyond the SM \cite{Chua:1998dx}, usually in supersymmetric extensions of it. Values of
$\mathcal{A}_{X_{s}\gamma}$ of up to $\pm 16\%$ were shown to be possible in specific models, while complying with stringent constraints from
electric dipole moments (EDMs). Including the long-distance contributions, it was shown in Ref.~\cite{Benzke:2010tq} that the the SM prediction using eq.~(\ref{acpshortlong})
is enlarged to the range $-0.6\% < \mathcal{A}_{X_{s}\gamma} < 2.8\%$, and (using updated estimates of the $\Lambda_{ij}$ parameters)
 is further increased to $-1.9\% < \mathcal{A}_{X_{s}\gamma} < 3.3\%$ in 
Ref.~\cite{Gunawardana:2019gep}. This revised SM prediction has decreased the effectiveness of $\mathcal{A}_{X_{s}\gamma}$ as a probe of physics beyond the SM. Consequently, in 
Ref.~\cite{Benzke:2010tq}  the difference of CP asymmetries  for the charged and neutral $B$ mesons
$\Delta\mathcal{A}_{X_{s}\gamma}=\mathcal{A}^\pm_{X_{s}\gamma}-\mathcal{A}^0_{X_{s}\gamma}$ was proposed, which is given by:
\bea
\Delta\mathcal{A}_{X_{s}\gamma} \approx 4 \pi^2 \alpha_s \frac{\tilde{\Lambda}_{78}}{m_b} \text{Im} \frac{C_{8g}}{C_{7\gamma}}\,.
\eea
This formula is obtained from eq.~(\ref{acpshortlong}) in which only the terms with $e_{\text{spec}}$ do not cancel out. The SM prediction
is $\Delta\mathcal{A}_{X_{s}\gamma}=0$ (due to the the Wilson coefficients being real) and hence this observable is potentially a more effective probe of
new physics than $\mathcal{A}_{X_{s}\gamma}$. Note that  $\Delta\mathcal{A}_{X_{s}\gamma}$ depends on the product of a long-distance term
(${\tilde{\Lambda}_{78}}$, whose value is only known to within an order of magnitude) and two short-distance terms ($C_8$ and $C_7$).

An alternative observable is the untagged (fully inclusive) asymmetry given by 
\begin{align}
	A_{X_{s+d}\gamma}=&\ 
\frac{(A_{X_s\gamma}^0+r_{0\pm}A_{X_s\gamma}^{\pm})
+R_{ds}(A_{X_d\gamma}^0+r_{0\pm}A_{X_d\gamma}^{\pm})}
{(1+r_{0\pm})(1+R_{ds})}\,.
\end{align}
Here $R_{ds}$ is the ratio ${\rm BR}({\overline B}\to d\gamma)/{\rm BR}({\overline B}\to s\gamma)\approx |V_{td}/V_{ts}|^2$.
The parameter $r_{0\pm}$ is defined as the following ratio:
\begin{equation}
r_{0\pm}\equiv \frac{N_{X_s}^++N_{X_s}^-}{N_{X_s}^{\bar{0}}+N_{X_s}^{0}},\,
\end{equation}
where $N_{X_s}^+$ is the number of $B^+$ mesons that decay to $X_s\gamma$ etc.
Experimentally, $r_{0\pm}\approx 1.03$ \cite{Kou:2018nap} and in our numerical analysis we take $r_{0\pm}=1$.
In the fully inclusive measurement of BR($b\to s/d \gamma$) the asymmetry 
$\mathcal{A}_{CP} ({\overline B} \to X_{s+d} \gamma)$ is measured by counting the difference in the number of positively and negatively charged leptons
from the tagged (not signal) $B$ meson. The SM prediction of $\mathcal{A}_{CP} ({\overline B} \to X_{s+d} \gamma)$ is essentially 0 \cite{Soares:1991te,Kagan:1998bh} (up to tiny $m^2_s/m^2_b$ corrections), 
even with the long-distance contribution included. Hence this observable is a cleaner test of new physics than $\mathcal{A}_{X_s \gamma}$. The first studies of the magnitude of the untagged asymmetry
in the context of physics beyond the SM were in Ref.~\cite{Akeroyd:2001cy}, and the importance of this observable was emphasised in Ref.~\cite{Hurth:2003dk}.
In this work we will consider the above three direct CP asymmetries in the context of 3HDMs: i) $\mathcal{A}_{X_s \gamma}$, ii) $\mathcal{A}_{CP} ({\overline B} \to X_{s+d} \gamma)$,  iii) $\Delta\mathcal{A}_{X_{s}\gamma}$.

Measurements of all three asymmetries have been carried out, and the most recent BELLE and BABAR measurements are summarised in Tab.~\ref{bsy_measurement}.
In Tab.~\ref{bsy_measurement} the CP asymmetry ${\mathcal{A}}^{\rm tot}_{X_s \gamma}$ would have the same magnitude as the average 
$\overline{\mathcal{A}}=(\mathcal{A}^0_{X_{s}\gamma}+\mathcal{A}^\pm_{X_{s}\gamma})/2$ if the production
cross-sections of $B^+B^-$ and $B^0\overline{B^0}$ were the same.  The BELLE measurement  \cite{Watanuki:2018xxg} of $\overline{\mathcal{A}}=(0.91\pm 1.21\pm 0.13)\%$ differs only
slightly from the BELLE measurement of ${\mathcal{A}}^{\rm tot}_{X_s \gamma}$ in Tab.~\ref{bsy_measurement}.
The world averages are taken from Ref.~\cite{Tanabashi:2018oca}. The given averages for  ${\mathcal{A}}^{\rm tot}_{X_s \gamma}$ and $\Delta\mathcal{A}_{X_{s}\gamma}$ are obtained from the two displayed 
measurements in Tab.~\ref{bsy_measurement}, while the average for $\mathcal{A}_{CP} ({\overline B} \to X_{s+d} \gamma)$ also includes two earlier BABAR measurements and 
the CLEO measurement $(-7.9\pm 10.8\pm 2.2)\%$ \cite{Coan:2000pu}. 

\begin{table}[h]
\begin{center}
\begin{tabular}{|c||c|c|c|c|}
\hline
& BELLE &  BABAR  & World Average\\ \hline
${\mathcal{A}}^{\rm tot}_{X_s \gamma}$ &   $(1.44\pm 1.28\pm 0.11)\%$ \cite{Watanuki:2018xxg}
&  $(1.73\pm 1.93\pm 1.02)\%$ \cite{Lees:2014uoa}   & $1.5\%\pm 1.1\%$ \cite{Tanabashi:2018oca}\\
$\mathcal{A}_{CP} ({\overline B} \to X_{s+d} \gamma)$ 
&  $(2.2\pm 3.9 \pm 0.9)\%$ \cite{Pesantez:2015fza} & $(5.7\pm 6.0\pm 1.8)\%$ \cite{Lees:2012ufa}  & $1.0\%\pm 3.1\%$ \cite{Tanabashi:2018oca}\\
$\Delta\mathcal{A}_{X_{s}\gamma}$ &   $(3.69\pm 2.65\pm 0.76)\%$ \cite{Watanuki:2018xxg}&  $(5.0\pm 3.9\pm 1.5)\%$ \cite{Lees:2014uoa}   & $4.1\%\pm 2.3\%$ \cite{Tanabashi:2018oca}\\
\hline
\end{tabular}
\end{center}
\caption{Measurements (given as a percentage) of ${\mathcal{A}}^{\rm tot}_{X_s \gamma}$,  $\mathcal{A}_{CP} ({\overline B} \to X_{s+d} \gamma)$ and $\Delta\mathcal{A}_{X_{s}\gamma}$ at BELLE, BABAR, and the
world average.}
\label{bsy_measurement}
\end{table}

At BELLE II all three asymmetries will be measured with greater precision \cite{Kou:2018nap}. At present around 74 fb$^{-1}$ of integrated luminosity have been accumulated, which is about one tenth of the
integrated luminosity at the BELLE experiment, and about one sixth that at the BABAR experiment. By the end of the year 2021 about 1 ab$^{-1}$ is expected, and thus
measurements of $b\to s\gamma$ at BELLE II will then match (or better) in precision those at BELLE and BABAR.
For an integrated luminosity of 50 ab$^{-1}$ (which is expected to be obtained by the end of the BELLE II experiment in around the year 2030), the estimated precision for 
$\mathcal{A}^{\rm tot}_{X_s \gamma}$ is 0.19\%, for $\mathcal{A}_{CP} ({\overline B} \to X_{s+d} \gamma)$ is $0.48\%$ (leptonic tag) and $0.7\%$ (hadronic tag), and for $\Delta\mathcal{A}_{X_{s}\gamma}$
is 0.3\% (sum-of-exclusives) and 1.3\% (fully inclusive with hadronic tag, and so it measures a sum of $b\to s\gamma$ and $b\to d\gamma$).
 These numbers are summarised in Tab.~\ref{bsy_prospects}, together with the SM predictions.
Due to the SM prediction of $\mathcal{A}_{CP} ({\overline B} \to X_{s+d} \gamma)$ being essentially zero,
a central value of 2.5\% with 0.5\% error would constitute a $5\sigma$ signal of physics beyond the SM. For $\Delta\mathcal{A}^{\rm tot}_{X_{s}\gamma}$, whose prediction in the SM is also essentially zero,
a central value of 1.5\% with 0.3\% error would constitute a $5\sigma$ signal. Note that the current $2\sigma$ allowed range of $\mathcal{A}^{\rm tot}_{X_s \gamma}$ is $-0.7\% < \mathcal{A}^{\rm tot}_{X_s \gamma} < 3.7\%$
($-1.8\% < \mathcal{A}^{\rm tot}_{X_s \gamma} < 4.8\%$ at $3\sigma$).
Comparing this range with the SM prediction of $-1.9\% < \mathcal{A}^{\rm tot}_{X_s \gamma} < 3.3\%$ shows that it is less likely that the observable $\mathcal{A}^{\rm tot}_{X_s \gamma}$ alone could provide a clear signal of
physics beyond the SM, e.g. a future central value of above $4.3\%$ (which is outside the current $2\sigma$ range) with the expected of error 0.19\% would be required to give a $5\sigma$ discrepancy from the upper SM prediction of 3.3\%.

\begin{table}[h]
\begin{center}
\begin{tabular}{|c||c|c|c|c|}
\hline
& SM Prediction &  Leptonic tag  & Hadronic tag & Sum of exclusives\\ \hline
$\mathcal{A}^{\rm tot}_{X_s \gamma}$ &   $-1.9\% < \mathcal{A}_{X_s \gamma} < 3.3\%$
&  x  & x & 0.19\%\\
$\mathcal{A}_{CP} ({\overline B} \to X_{s+d} \gamma)$ 
&  0 & 0.48\% & 0.70\% & x\\
$\Delta\mathcal{A}_{X_{s}\gamma}$ & 0 &  x & 1.3\% &  0.3\%\\
\hline
\end{tabular}
\end{center}
\caption{SM predictions of $\mathcal{A}^{\rm tot}_{X_s \gamma}$,  $\mathcal{A}_{CP} ({\overline B} \to X_{s+d} \gamma)$ and $\Delta\mathcal{A}_{X_{s}\gamma}$, and expected experimental errors in their measurements 
at BELLE II with 50 ab$^{-1}$.}
\label{bsy_prospects}
\end{table}

\section{The decays $\overline{B} \to X_s \gamma$ and ${\overline B} \to X_{s+d} \gamma$ in the 3HDM}
\noindent
In this section the theoretical structure of the 3HDM is briefly introduced, followed by a discussion of the
Wilson coefficients. Finally, the expressions for the BRs of ${\overline B} \to X_s \gamma$ and ${\overline B} \to X_{d} \gamma$ are given.

\subsection{Fermionic couplings of the charged scalars in a  3HDM}
\noindent
In a 3HDM, two $SU(2)\otimes U(1)$ isospin scalar doublets (with hypercharge $Y=1$) are added to the Lagrangian of the SM.
There are two (physical) charged scalars and for a more detailed description of the
model we refer the reader to Refs.~\cite{Cree:2011uy, Akeroyd:2019mvt}.
In order to eliminate tree-level flavour changing neutral
currents (FCNCs) that are mediated by scalars, the couplings of the scalar doublets to fermions ("Yukawa couplings") are assumed to be invariant under certain discrete symmetries
(a requirement called "natural flavour conservation" (NFC), e.g. see Refs.~\cite{Glashow:1976nt,Branco:2011iw}).
The Lagrangian that describes the interactions of $H_1^\pm$ and $H^\pm_2$ (the two charged scalars of the 3HDM, which we do not order in mass) with 
the fermions is given as follows:
\begin{eqnarray}
{\cal L}_{H^\pm} =
-\left\{ \frac{\sqrt2V_{ud}}{v}\overline{u}
\left(m_d X_1{P}_R+m_u Y_1{P}_L\right)d\,H^+_1
+\frac{\sqrt2m_\ell }{v} Z_1\overline{\nu_L^{}}\ell_R^{}H^+_1
+{H.c.}\right\}  \nonumber \\
+ \left\{ \frac{\sqrt2V_{ud}}{v}\overline{u}
\left(m_d X_2{P}_R+m_u Y_2{P}_L\right)d\,H^+_2
+\frac{\sqrt2m_\ell }{v} Z_2\overline{\nu_L^{}}\ell_R^{}H^+_2
+{H.c.} \right\}\,.
\label{lagrangian}
\end{eqnarray}
Here $u(d)$ refers to the up(down)-type quarks, and $\ell$
refers to the electron, muon and tau.
 In a 2HDM there is only
one charged scalar, and the parameters $X$, $Y$, and $Z$ (with no subscript) are equal to $\tan\beta$ or $\cot\beta$ (where $\tan\beta=v_2/v_1$, the ratio of vacuum expectation values).
In contrast, in a 3HDM the $X_i$, $Y_i$, and $Z_i$ ($i=1,2$) each depend on four parameters
of a unitary matrix $U$, and thus the phenomenology of $H^\pm_1$ and $H^\pm_2$ can differ from that of $H^\pm$ in a 2HDM.
This matrix $U$ connects the charged scalar fields in the 
weak eigenbasis ($\phi^\pm_1,\phi^\pm_2,\phi^\pm_3)$ with the 
physical scalar fields
($H^\pm_1$, $H^\pm_2$) and the charged Goldstone boson $G^\pm$ as follows:
\begin{equation}
	\left( \begin{array}{c} G^+ \\ H_1^+ \\ H_2^+ \end{array} \right) 
	= U \left( \begin{array}{c} \phi_1^+ \\ \phi_2^+ \\ \phi_3^+ \end{array} \right).
	\label{eq:Udef}
\end{equation}
The couplings $X_i$, $Y_i$ and $Z_i$ in terms of the elements of $U$ are
\cite{Cree:2011uy}: 
\begin{equation}
	X_1 = \frac{U_{d2}^\dagger}{U_{d1}^\dagger}, \quad \quad 
	Y_1 = - \frac{U_{u2}^\dagger}{U_{u1}^\dagger}, \quad \quad 
	Z_1 = \frac{U_{\ell 2}^\dagger}{U_{\ell 1}^\dagger}\,,
\label{eq:xyz}
\end{equation}
and 
\begin{equation}
	X_2 = \frac{U_{d3}^\dagger}{U_{d1}^\dagger}, \quad \quad 
	Y_2 = - \frac{U_{u3}^\dagger}{U_{u1}^\dagger}, \quad \quad 
	Z_2 = \frac{U_{\ell 3}^\dagger}{U_{\ell 1}^\dagger}\,.
\label{eq:xyz}
\end{equation}

The values of $d$, $u$, and $\ell$ in these matrix elements  are given in Tab.~\ref{valuesudl} and 
depend on which of the five distinct 3HDMs is being considered. 
The choice of  $d=1$, $u=2$, and $\ell=3$ indicates that the down-type quarks receive their mass from $v_1$,
the up-type quarks from $v_2$, and the charged leptons from $v_3$ (and is called the ``Democratic 3HDM''). The other
possible choices of $d$, $u$, and $\ell$ in a 3HDM are given the same names as the four types of 2HDM.
\begin{table}[h]
\begin{center}
\begin{tabular}{|c||c|c|c|}
\hline
& $u$ &  $d$ &  $\ell$ \\ \hline
3HDM (Type I) &  2 & 2 & 2 \\
3HDM (Type II) & 2 & 1 & 1 \\
3HDM (Lepton-specific) & 2 & 2 & 1 \\
3HDM (Flipped) & 2 & 1 & 2 \\
3HDM (Democratic) & 2 & 1 & 3 \\
\hline
\end{tabular}
\end{center}
\caption{The five versions of the 3HDM with NFC,
and the corresponding values of $u$, $d$, and $\ell$. The choice of $u=2$ means that the up-type quarks receive their mass
from the VEV $v_2$, and likewise for $d$ (down-type quarks) and $\ell$ (charged leptons).}
\label{valuesudl}
\end{table}

The elements of the matrix $U$ can be parametrised by
four parameters  $\tan\beta$, $\tan\gamma$, $\theta$, and $\delta$, where
\begin{equation}
	\tan\beta = v_2/v_1, \qquad \tan\gamma = \sqrt{v_1^2 + v_2^2}/v_{3}\,.
\end{equation}
The angle $\theta$ and phase $\delta$ can be written explicitly as  
functions of several parameters in the scalar potential \cite{Cree:2011uy}.
The explicit form of $U$ is:
\begin{eqnarray}
	U &=& \left( \begin{array}{ccc} 
		1 & 0 & 0 \\
		0 & e^{-i \delta} & 0 \\
		0 & 0 & 1 \end{array} \right)
		\left( \begin{array}{ccc}
		1 & 0 & 0 \\
		0 & c_\theta & s_\theta e^{i \delta} \\
		0 & -s_\theta e^{-i \delta} & c_\theta \end{array} \right)
		\left( \begin{array}{ccc}
		s_\gamma & 0 & c_\gamma \\
		0 & 1 & 0 \\
		-c_\gamma & 0 & s_\gamma \end{array} \right)
		\left( \begin{array}{ccc}
		c_\beta & s_\beta & 0 \\
		-s_\beta & c_\beta & 0 \\
		0 & 0 & 1 \end{array} \right)
	\nonumber \\
	&=& \left( \begin{array}{ccc}
	s_\gamma c_\beta & s_\gamma s_\beta 	& c_\gamma \\
	-c_\theta s_\beta e^{-i\delta} - s_\theta c_\gamma c_\beta 
		& c_\theta c_\beta e^{-i\delta} - s_\theta c_\gamma s_\beta & s_\theta s_\gamma \\
	s_\theta s_\beta e^{-i\delta} - c_\theta c_\gamma c_\beta 
		& -s_\theta c_\beta e^{-i\delta} - c_\theta c_\gamma s_\beta & c_\theta s_\gamma 
	\end{array} \right).
	\label{eq:Uexplicit}
\end{eqnarray}
Here $s$ and $c$ denote the sine or cosine of the respective angle. Hence the functional forms of
$X_i$, $Y_i$, and $Z_i$ in a 3HDM depend on four parameters. As mentioned earlier, this is in contrast to the analogous couplings in 
the 2HDM for which $\tan\beta$ is the only free coupling parameter.

The parameters $X_i$, $Y_i$ and $Z_i$ are constrained (for a specific value of $m_{H_i^\pm}$) by direct searches for $H^\pm_i$ (e.g. at the LHC) and 
by their effect on low-energy observables in flavour physics. A summary of these constraints can be found in Ref.~\cite{Cree:2011uy}, in which the lightest charged scalar is assumed to give the
dominant contribution to the observable being considered. A full study in the context of the 3HDM with both charged scalars contributing significantly has not been performed, and
is beyond the scope of this work.
The coupling $Y_i$ is most strongly constrained from the process $Z\to b\overline b$ 
from LEP data. For $m_{H^\pm}$ around 100 GeV 
the constraint is roughly $|Y_i|<0.8$ (assuming $|X_i|\le 50$, so that the
dominant contribution is from the $Y_i$ coupling), and weakens with increasing mass of the charged scalar.
Constraints on the $X_i$ and $Z_i$ are weaker and we take $|X_i|<50$ and $|Z_i|<50$ as being representative of these constraints for $m_{H_i^\pm}$ around 100 GeV.

The couplings $Z_i$ do not enter the partial width of $b\to s\gamma$, and only
the couplings to quarks are relevant ($X_i$ and $Y_i$). 
Consequently, the partial width for $b\to s\gamma$ in Type I and the lepton-specific
structures (which have identical functional forms for $X_i$ and $Y_i$ due to $u=d$ in Tab.~\ref{valuesudl}) has the same dependence on the parameters
of $U$. Likewise, the partial width for $b\to s\gamma$  in 
Type II, flipped and democratic structures ($u\ne d$ in Tab.~\ref{valuesudl}) is the same.  The contribution of $H^\pm_1$  and $H^\pm_2$ to
BR({${\overline B} \to X_s  \gamma$}) has been studied in the 3HDM at the leading order (LO) in Ref.~\cite{Hewett:1994bd} (no $\alpha_s$ corrections arising from diagrams with charged scalars)
and at next-to-leading order (NLO) in Ref.~\cite{Akeroyd:2016ssd} ($\alpha_s$ corrections arising from diagrams with charged scalars). In Ref.~\cite{Akeroyd:2016ssd} the effect of a non-zero phase $\delta$
was not studied, and direct CP asymmetries were not considered. Previous studies of $\mathcal{A}_{X_s \gamma}$ (and $\mathcal{A}_{X_d \gamma}$),
but not $\mathcal{A}_{CP} ({\overline B} \to X_{s+d} \gamma)$ and $\Delta\mathcal{A}_{X_{s}\gamma}$, in models with one charged scalar (e.g. 2HDM, or the lightest $H^\pm$ of a 3HDM or multi-Higgs doublet model)
have been carried out in Refs.~\cite{Wolfenstein:1994jw,Asatrian:1996as,Borzumati:1998tg,Kiers:2000xy,Jung:2010ab,Jung:2012vu}.

\subsection{Wilson coefficients in 3HDM}
\noindent
The direct CP asymmetry given by eq.~(\ref{acpshortlong}) is written in terms of Wilson coefficients, which (for $B$ observables) are generally evaluated at the scale of $\mu_b=m_b$. We use the explicit formulae in 
Ref.~\cite{Borzumati:1998tg} for the Wilson coefficients at LO and NLO in the 2HDM, and apply them to the 3HDM (generalising the expressions to account for the two charged scalars).
The LO Wilson coefficients \cite{Hewett:1994bd} at the matching scale $\mu_W=m_W$ are as follows:
\begin{align}
C^{0,\text{eff}}_2 (\mu_W) &= 1,\\
C^{0,\text{eff}}_i (\mu_W) &= 0 \quad (i = 1,3,4,5,6) \\
C^{0,\text{eff}}_{7\gamma} (\mu_W) &= C^{0}_{7,SM}  + |Y_1 |^2 C^{0}_{7,Y_1Y_1} + |Y_2 |^2 C^{0}_{7,Y_2Y_2}  
+ (X_1^{}Y_1^*) C^{0}_{7,X_1Y_1} + (X_2^{}Y_2^*) C^{0}_{7,X_2Y_2} \\
C^{0,\text{eff}}_{8g} (\mu_W) &= C^{0}_{8,SM}  + |Y_1 |^2 C^{0}_{8,Y_1Y_1} + |Y_2 |^2 C^{0}_{8,Y_2Y_2}  
+ (X_1Y_1^*) C^{0}_{8,X_1^{}Y_1} + (X_2^{}Y_2^*) C^{0}_{8,X_2Y_2} \,.
\end{align}
Terms with $X^*_1Y_1$, $X^*_2Y_2$, $|X_1|^2$ and $|X_2|^2$ are absent because $m_s=0$ (as is usually taken) in the effective Hamiltonian.
Explicit forms for all $C^0_{7}$ and $C^0_{8}$ are given in Ref.~\cite{Borzumati:1998tg}: those for the SM contribution are functions of $m_t^2/m^2_W$ while those for
$H^\pm_1$ and $H^\pm_2$ are functions of  $m_t^2/m^2_{H^\pm_1}$ and $m_t^2/m^2_{H^\pm_2}$, respectively.

The NLO Wilson coefficients at the matching scale are as follows:
\begin{align}
C^{1,\text{eff}}_1 (\mu_W) =&\   15 + 6 \hspace{0.2cm}  \text{ln} \frac{\mu^2_{W}}{M^2_W} ,\\
C^{1,\text{eff}}_4 (\mu_W) =&\ E_0 + \frac{2}{3}  \hspace{0.2cm}  \text{ln}  \frac{\mu^2_{W}}{M^2_W} + |Y_1|^2 E_{H_2} + |Y_2 |^2 E_{H_3} \\
C^{1,\text{eff}}_i (\mu_W) =&\  0 \quad (i = 2,3,5,6) \\
C^{1,\text{eff}}_{7\gamma} (\mu_W) =&\  C^{1,\text{eff}}_{7,SM}(\mu_W)  + |Y_1 |^2 C^{1,\text{eff}}_{7,Y_1Y_1}(\mu_W)  
+|Y_2 |^2 C^{1,\text{eff}}_{7,Y_2Y_2}(\mu_W) 
 \nonumber \\
&\ + (X_1^{}Y_1^*) C^{1,\text{eff}}_{7,X_1Y_1}(\mu_W) + (X_2^{}Y_2^*) C^{1,\text{eff}}_{7,X_2Y_2}(\mu_W) \\
C^{1,\text{eff}}_{8g} (\mu_W) =&\ C^{1,\text{eff}}_{8,SM}(\mu_W)  + |Y_1 |^2 C^{1,\text{eff}}_{8,Y_1Y_1}(\mu_W)
+|Y_2 |^2 C^{1,\text{eff}}_{8,Y_2Y_2}(\mu_W) \nonumber  \\  
&\ + (X_1^{}Y_1^*) C^{1,\text{eff}}_{8,X_1Y_1}(\mu_W) + (X_2^{}Y_2^*) C^{1,\text{eff}}_{8,X_2Y_2}(\mu_W) \,. 
\end{align}

Explicit forms for all functions are given in Ref.~\cite{Borzumati:1998tg}. Renormalisation group running is used to evaluate the Wilson coefficients at the scale $\mu=m_b$.

The partial width for ${\overline B} \to X_s \gamma$ has four distinct parts: i) Short-distance contribution from
the $b\to s\gamma$ partonic process (to a given order in perturbation theory); ii) Short-distance contribution from the $b\to s\gamma g$ partonic process; iii) and iv) Non-perturbative corrections that scale as
$1/m_b^2$ and $1/m_c^2$, respectively. The partial width of $\bar{B} \to X_s \gamma$ is as follows:
\bea
\Gamma ({\overline B} \to X_s \gamma) &=& \frac{G^2_F}{32\pi^4} |V^{*}_{ts} V_{tb} |^2 \alpha_{em} m^5_b    \nonumber \\&\times &
\Bigg \{ |\bar{D} |^2 + A + \frac{\delta^{NP}_{\gamma}}{m^2_b} |\text{C}^{0,\text{eff}}_{7} (\mu_b) |^2  \nonumber \\ &+& \frac{\delta^{NP}_{c}}{m^2_c} {\rm Re} \Bigg[  [\text{C}^{0,\text{eff}}_{7} (\mu_b)]^* \times \bigg ( \text{C}^{0,\text{eff}}_{2} (\mu_b) - \frac{1}{6} \text{C}^{0,\text{eff}}_{1} (\mu_b)\bigg) \Bigg] \Bigg\} \label{pwbtosg}\,.
\eea 
The short-distance contribution is contained in  $|\bar{D} |^2$, with  $\bar{D}$ given by:
\begin{equation}
\bar{D} =  \text{C}^{0,\text{eff}}_{7} (\mu_b) + \frac{\alpha_s (\mu_b)}{4 \pi} [  \text{C}^{1,\text{eff}}_{7} (\mu_b) + V(\mu_b)]\,.
\end{equation}
The Wilson coefficient $C^{0,\text{eff}}_{7} (\mu_b)$ is a linear combination of $C^{0,\text{eff}}_{7} (\mu_W)$, $C^{0,\text{eff}}_{8} (\mu_W)$ and $C^{0,\text{eff}}_{2} (\mu_W)$, while
$C^{1,\text{eff}}_{7} (\mu_b)$ is a linear combination of these three LO coefficients as well as the NLO coefficients $C^{1,\text{eff}}_{7} (\mu_W)$, $C^{1,\text{eff}}_{8} (\mu_W)$, $C^{1,\text{eff}}_{4} (\mu_W)$, and $C^{1,\text{eff}}_{1} (\mu_W)$.
The parameter $V(\mu_b)$ is a summation over all the LO Wilson coefficients which are evaluated at the scale $\mu_b=m_b$.  The contribution from $b\to s\gamma g$ is contained in $A$, and the remaining two terms are the non-perturbative contributions.
\noindent
In $|\bar{D} |^2$ there are terms of order $\alpha_s^2$, but to only keep terms to the NLO order for a consistent calculation (to  $\alpha_s$) the following form is used in Ref.~\cite{Borzumati:1998tg}:
\be
|\bar{D} |^2 = |C^{0,\text{eff}}_{7} (\mu_b)|^2 \{ 1 + 2  {\rm Re}( \Delta \bar{D})\}\,,
\ed
\be
\Delta \bar{D} =  \frac{\bar{D} - C^{0,\text{eff}}_{7} (\mu_b) }{C^{0,\text{eff}}_{7} (\mu_b)} =  \frac{\alpha_s (\mu_b)}{4 \pi} \frac{C^{1,\text{eff}}_{7} (\mu_b) + V(\mu_b)}{C^{0,\text{eff}}_{7} (\mu_b)}\,.
\ed
The $m^5_b$ dependence is removed by using the measured value of the semi-leptonic branching ratio
${\rm BR}_{SL}\approx 0.1$ and its partial width $\Gamma_{SL}$ (which also depends on $m^5_b$),  and
BR$({\overline B} \to X_s \gamma)$ can be written as follows:
\begin{equation}
{\rm BR}({\overline B} \to X_s \gamma) = \frac{\Gamma ({\overline B} \to X_s \gamma)}{\Gamma_{SL}} {\rm BR}_{SL}\,.
\end{equation}

\section{Numerical results}
\noindent
The four input parameters that determine $X_i$, $Y_i$, and $Z_i$ 
are varied in the following ranges, while respecting the constraints  $|X_i|<50$, $|Z_i|<50$ and $|Y_i|<0.8$ for $m_{H^\pm_i}=100$ GeV.
\begin{eqnarray}
-\pi/2\le \theta \le 0,\;\;\;\;\;\;\;\;\;\;  0\le \delta \le 2\pi,   
\nonumber \\
0.1 \le \tan\beta \le 60, \;\;\;\;\; 0.1 \le \tan\gamma \le 60 \,.
\label{4param}
\end{eqnarray}
As mentioned in section III.A, the functional dependence on these four input parameters of the observables BR$(b\to s\gamma)$, ${\mathcal{A}}_{X_s \gamma}$,  $\mathcal{A}_{CP} ({\overline B} \to X_{s+d} \gamma)$ and 
$\Delta\mathcal{A}_{X_{s}\gamma}$ is the 
same in the Flipped 3HDM, Type II and Democratic 3HDMs. Results will be shown in this class of models, and sizeable values of the asymmetries are shown to be possible.
Results are not shown for the Model Type I and lepton specific structures because the asymmetries in these two models do not differ much from the SM values, the reason being that
the products $X_1Y_1^*$ and $X_2Y_2^*$ (which enter the Wilson coefficients) are real in these two models, leading to real $C_7$ and $C_8$.
The couplings $Z_i$ are different functions of $\theta$, $\tan\beta$, $\tan\gamma$ and $\delta$ in the Flipped 3HDM, Type II and Democratic 3HDMs, and thus the constraints in
eq.~(\ref{4param}) on $Z_i$ rule out different regions of the four input parameters in each model. However, the constraints from $Z_i\le 50$ are quite weak, and so the allowed parameter space 
from $|X_i|<50$, $|Z_i|<50$ and $|Y_i|<0.8$ for $m_{H^\pm_i}=100$ GeV is essentially the same in all three models under consideration. For definiteness, our
results will presented in the context of the Flipped 3HDM. In eq.~(\ref{bsy_exp}) for the measurement of BR$({\overline B}\to X_s\gamma)$ we take the $3\sigma$ allowed range, 
giving $2.87\times 10^{-4}\le {\rm BR}({\overline B}\to X_s\gamma)\le 3.77\times 10^{-4}$. 
\begin{figure}[!b]
    \centering
     \begin{subfigure}
    {
        \includegraphics[scale=.5]{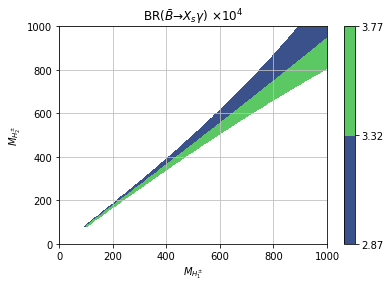}
    }
    {
        \includegraphics[scale=.5]{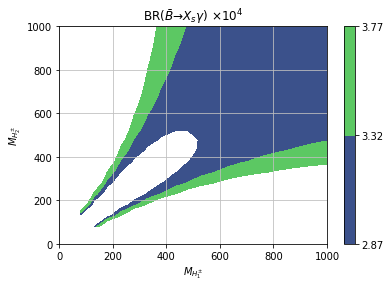}
    }
    \caption{BR$({\overline B}\to X_s\gamma)$  in the plane $[m_{H^\pm_1}, m_{H^\pm_2}$], with $\theta=-\pi/4$, $\tan\beta=10$, $\tan\gamma=1$.
    Left Panel: $\delta=0$.  Right Panel: $\delta=\pi/2$.} \label{BRdel0}
     \subfigure
    {
        \includegraphics[scale=.5]{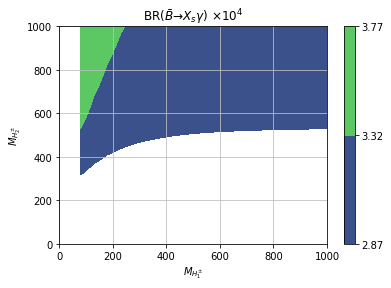}
    }
      \subfigure
    {
        \includegraphics[scale=.5]{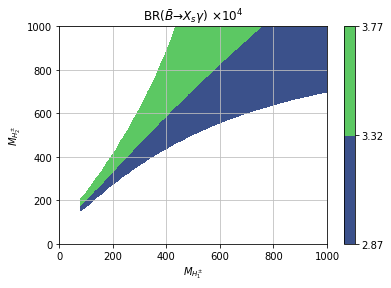}
    }
    \end{subfigure}
    \caption
    {BR$({\overline B}\to X_s\gamma)$  in the plane $[m_{H^\pm_1}, m_{H^\pm_2}$], with $\theta=-\pi/2.1$, $\tan\beta=10$, $\tan\gamma=1$.
    Left Panel: $\delta=0$.  Right Panel: $\delta=\pi/2.$
    } \label{brcbcs80}
\end{figure}

\begin{figure}[!b]
    \centering
     \begin{subfigure}
    {
        \includegraphics[scale=.5]{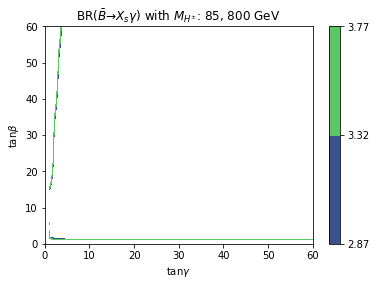}
    }
    {
        \includegraphics[scale=.5]{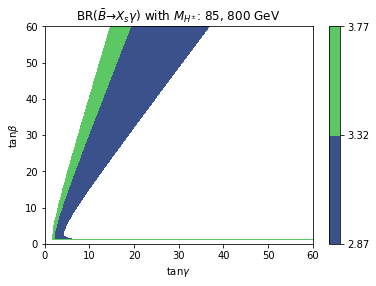}
    }
    \caption{BR$({\overline B}\to X_s\gamma)$  in the plane $[\tan\gamma, \tan\beta $], with 
    $\theta=-\pi/3$,
 $m_{H^\pm_1}=$85 GeV, $m_{H^\pm_2}=800$ GeV.
    Left Panel: $\delta=0$.  Right Panel: $\delta=\pi/2$.}  \label{br80}
     \subfigure
    {
        \includegraphics[scale=.5]{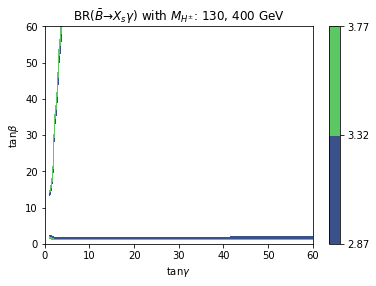}
    }
      \subfigure
    {
        \includegraphics[scale=.5]{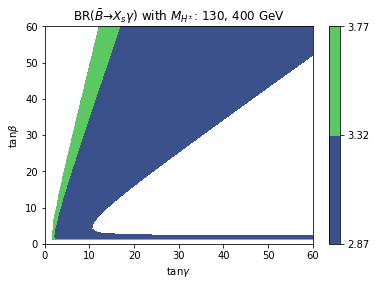}
    }
    \end{subfigure}
    \caption
    {BR$({\overline B}\to X_s\gamma)$  in the plane $[\tan\gamma, \tan\beta $], with 
    $\theta=-\pi/3$,
 $m_{H^\pm_1}=$130 GeV, $m_{H^\pm_2}=400$ GeV.
    Left Panel: $\delta=\pi/4$.  Right Panel: $\delta=\pi/2$.}   \label{brcbcs80}
\end{figure}

In Figs.~1a and 1b the magnitude of BR$(b\to s\gamma)$ in the plane $[m_{H^\pm_1}, m_{H^\pm_2}]$ is plotted with
$\theta=-\pi/4$, $\tan\beta=10$ and $\tan\gamma=1$. In the left panel $\delta=0$ and in the right panel $\delta=\pi/2$. 
In Ref.~\cite{Akeroyd:2016ssd} only $\delta=0$ was taken when studying BR$(b\to s\gamma)$ in 3HDMs. 
In our numerical analysis we set the normalisation scale to be $\mu_b=m_b=4.77$ GeV (the central value of the $b-$quark pole mass), and the uncertainty in the asymmetries due to the choice of $\mu_b$ is discussed later. 
It can be seen in Fig.~1a and Fig.~1b that for this choice of parameters 
the non-zero value of $\delta$ significantly increases the allowed parameter space in the plane $[m_{H^\pm_1}, m_{H^\pm_2}]$.
In Figs.~2a and 2b the parameters are taken to be $\theta=-\pi/2.1$, $\tan\beta=10$ and $\tan\gamma=1$. In the left panel $\delta=0$ and in the right panel $\delta=\pi/2$.
In this case the non-zero value of $\delta$ significantly decreases the parameter space in the plane $[m_{H^\pm_1}, m_{H^\pm_2}]$, although a region with $m_{H^\pm_1}<m_t$ and $m_{H^\pm_2}<m_t$ 
becomes allowed for $\delta=\pi/2$.
In all these plots the notation $m_{H_1^\pm}>m_{H^\pm_2}$ is not used and both 
masses are scanned in the range 80 GeV$<  m_{H_1^\pm},  m_{H^\pm_2}<1000$ GeV. It is clear that the phase $\delta$ can have a sizeable effect the parameter
space of $[m_{H^\pm_1}, m_{H^\pm_2}]$ in the 3HDM.  

In an earlier work by some of us \cite{Akeroyd:2018axd} the region allowed by BR$(b\to s\gamma)$ in the plane $[\tan\gamma,\tan\beta]$ in the Flipped 3HDM was obtained by using the 
constraint $-0.7 < {\rm Re}(X_1Y^*_1 )< 1.1$ only, with $\delta=0$. This
 is a result from the
Aligned 2HDM for small $|Y_1|^2$, and when applied to an $H^\pm$ of the 3HDM it is neglecting the contributions of $X_2Y_2^*$,  $|Y^2_2|$ and $m_{H^\pm_2}$. In 
Fig.~3a and Fig.~3b we compare this approximation with the full BR$(b\to s\gamma)$ constraint in the 3HDM.
In Fig.~3a, the allowed region in the plane $[\tan\gamma,\tan\beta]$ is plotted with $\theta=-\pi/3$,
 $m_{H^\pm_1}=$85 GeV, $m_{H^\pm_2}=800$ GeV with $\delta=0$. The region is much smaller than that displayed in Ref.~\cite{Akeroyd:2018axd}, which used the constraint $-0.7 < {\rm Re}(X_1Y^*_1 )< 1.1$ in the same plane;
 decreasing $m_{H^\pm_2}$ below 600 GeV leads to no allowed parameter space of  $[\tan\gamma,\tan\beta]$ for this choice of parameters.
In Fig.~3b, which has $\delta=\pi/2$, but other parameters the same as in Fig.~3a, one can see that the allowed region is much larger, and is in fact more similar in extent (although still smaller) than that allowed from  
the constraint $-0.7 < {\rm Re}(X_1Y^*_1 )< 1.1$ with $\delta=0$ in Ref.~\cite{Akeroyd:2018axd}. Hence the approximate constraint does not give a very accurate exclusion of parameter space, but the inclusion of a non-zero value of $\delta$ can 
(very roughly) reproduce the allowed regions in 
Ref.~\cite{Akeroyd:2018axd} (which focussed on the possibility of a large BR$(H^\pm\to cb$) in the Flipped 3HDM with $\delta=0$). In Fig.~4a and Fig.~4b we take $m_{H^\pm_1}=$130 GeV, $m_{H^\pm_2}=400$ GeV
(i.e. a smaller mass splitting between the charged scalars) and $\theta=-\pi/3$. In Fig.~4a we take $\delta=\pi/4$  and in Fig.~4b $\delta=\pi/2$.
One can see
that for $\delta=\pi/4$ very little parameter space is allowed by BR$(b\to s\gamma)$. In contrast, for $\delta=\pi/2$ a sizeable region of the plane $[\tan\gamma,\tan\beta]$ is permitted.
We calculated BR$(H^\pm\to cb$) in the same plane $[\tan\gamma,\tan\beta]$ but with $\delta=\pi/2$ and found that is essentially the same as the case with $\delta=0$ in 
Ref.~\cite{Akeroyd:2018axd}. Hence there is a sizeable parameter space for a large BR$(H^\pm\to cb$) in the Flipped 3HDM while satisfying the full BR$(b\to s\gamma)$ constraint, provided that 
$\delta$ is non-zero.
 
 We now turn our attention to the CP asymmetries. For ${\mathcal{A}}_{X_s \gamma}$ we use $\overline{\mathcal{A}}=(\mathcal{A}^0_{X_{s}\gamma}+\mathcal{A}^\pm_{X_{s}\gamma})/2$,
which is obtained by taking $e_{\text{spec}}=1/6$.
The CP asymmetries are evaluated at $\mathcal{O}(\alpha_s)$, so that we use the LO formulae for the Wilson coefficients $C_2$, $C_{7\gamma}$, and $C_{8g}$ in eq.~(\ref{acpshortlong}). 
In order to evaluate the CP asymmetries at $\mathcal{O}(\alpha_s^2)$, it is necessary to include not only the NLO terms of these Wilson coefficients but also the NNLO terms of $C_{7\gamma}$ and $C_{8g}$. 

In Fig.~5a, Fig.~5b, and Fig.~6 the asymmetries ${\mathcal{A}}_{X_s \gamma}$, $\Delta A_{CP}$, and $\mathcal{A}_{CP} ({\overline B} \to X_{s+d} \gamma)$ are (respectively) 
plotted in the plane $[\tan\gamma,\tan\beta]$. In all these figures the remaining four 3HDM parameters are fixed as $m_{H^{\pm}_1} = 170$ GeV, $m_{H^{\pm}_2} = 180$ GeV, $\theta=-\pi/4$ and
  $\delta=2.64$, whereas the long-distance (hadronic) parameters are taken to be $\tilde{\Lambda}^{u}_{17} = 0.66$ GeV, $\tilde{\Lambda}^{c}_{17} = 0.010$ GeV and $\tilde{\Lambda}_{78} = 0.19$ GeV.
 \begin{figure}[!h]
    \centering
     \begin{subfigure}
    {
        \includegraphics[scale=.5]{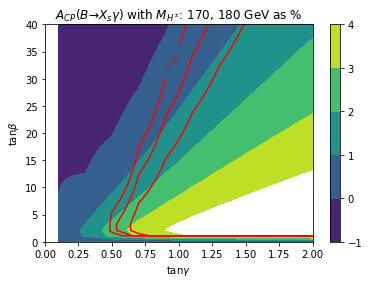}
    }
    {
        \includegraphics[scale=.5]{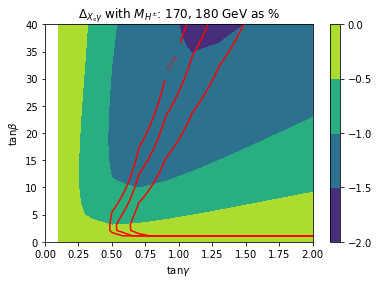}
    }
    \caption{CP asymmetries (as a percentage) in the plane [$\tan\gamma, \tan\beta$] with $m_{H^{\pm}_1} = 170$ GeV, $m_{H^{\pm}_2} = 180$ GeV, $\theta=-\pi/4$ and
    $\delta=2.64$.  The three red lines (from left to right) show the upper ($3\sigma$) limit, the central value, and the lower ($3\sigma$) limit for BR$({\overline B}\to X_s\gamma)$.
    Left Panel: $\mathcal{A}_{CP} ({\overline B} \to X_s \gamma )$, with the white region for $\tan\gamma>1$ violating the $3\sigma$ experimental bounds.
    Right Panel:  $\Delta A_{CP}$. }  
     \subfigure
    {
        \includegraphics[scale=.5]{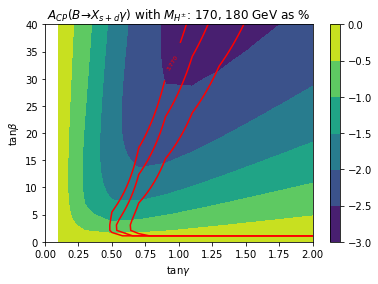}
    }
    \end{subfigure}
    \caption
    { $\mathcal{A}_{CP} ({\overline B} \to X_{s+d} \gamma)$ (as a percentage) in the plane [$\tan\gamma, \tan\beta$] with $m_{H^{\pm}_1} = 170$ GeV, $m_{H^{\pm}_2} = 180$ GeV, $\theta=-\pi/4$ and
    $\delta=2.64$.  The three red lines (from left to right) show the upper ($3\sigma$) limit, the central value, and the lower ($3\sigma$) limit for BR$({\overline B}\to X_s\gamma)$.
    } \label{brcbcs80}
\end{figure}
  The scale $\mu_b$ is taken to be 4.77 GeV (pole mass $m_b$).
 The three red lines (from left to right) show the upper ($3\sigma$) limit, the central value, and the lower ($3\sigma$) limit for BR$({\overline B}\to X_s\gamma)$.
 The white region in Fig.5a with roughly $\tan\gamma>1$ violates the current experimental ($3\sigma$) limit for  ${\mathcal{A}}_{X_s \gamma}$ (the white regions in
 Figs.5a, 5b and 6 with $\tan\gamma<0.1$ correspond to parameter choices not covered in the scan).
In Fig.~5a, in the parameter space allowed by BR$({\overline B}\to X_s\gamma)$  the magnitude of  ${\mathcal{A}}_{X_s \gamma}$ is roughly between $0.5\%$ and $1.5\%$, which is within the current experimental limits.
In Fig.~5b, $\Delta A_{CP}$ can reach $-1.5\%$, which would provide a $5\sigma$ signal at BELLE II with 50 ab${^{-1}}$. We note that $\Delta A_{CP}$ is directly proportional to 
$\tilde{\Lambda}_{78}$, which has been taken to have its largest allowed value. If $\tilde{\Lambda}_{78}$ is reduced then $\Delta A_{CP}$ will decrease proportionally.
In Fig.6 it is shown that $\mathcal{A}_{CP} ({\overline B} \to X_{s+d} \gamma)$ can reach almost $-3\%$, which would be a $5\sigma$ signal at BELLE II. The 
parameter $\tilde{\Lambda}_{78}$ has a subdominant effect on $\mathcal{A}_{CP} ({\overline B} \to X_{s+d} \gamma)$ (in contrast to  $\Delta A_{CP}$) and so $\mathcal{A}_{CP} ({\overline B} \to X_{s+d} \gamma)\approx -3\%$ is possible, independent of the
value of  $\tilde{\Lambda}_{78}$. We note that there is more parameter space in a 3HDM for such large asymmetries than in the Aligned 2HDM \cite{Jung:2010ab,Jung:2012vu}. This is because there is more possibility for
cancellation in the contributions of $H^\pm_1$ and $H^\pm_2$ to ${\overline B}\to X_s\gamma$ (while having a large asymmetry), but in the Aligned 2HDM there is only one charged scalar and 
no $X_2$ and $Y_2$ coupling.
\begin{figure}[!h]
    \centering
     \begin{subfigure}
    {
        \includegraphics[scale=.5]{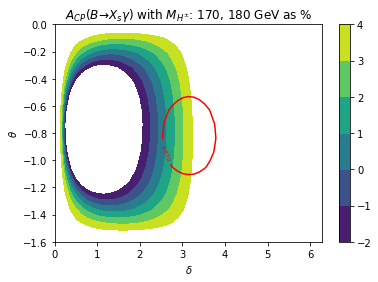}
    }
    {
        \includegraphics[scale=.5]{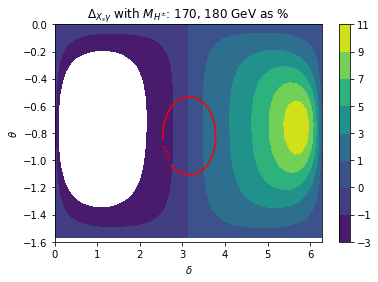}
    }
    \caption{CP asymmetries (as a percentage) in the plane [$\delta, \theta$] with $m_{H^{\pm}_1} = 170$ GeV, $m_{H^{\pm}_2} = 180$ GeV, $\tan\beta =35$ and
    $\tan\gamma =1.32$. Inside the red circles the predicted BR$({\overline B}\to X_s\gamma)$ satisfies 
the current experimental constraint. The white regions are excluded by the current ($3\sigma$) experimental limits on the asymmetry displayed in the figure.
    Left Panel: $\mathcal{A}_{CP} ({\overline B} \to X_s \gamma )$. 
    Right Panel:  $\Delta A_{CP}$. }  
     \subfigure
    {
        \includegraphics[scale=.5]{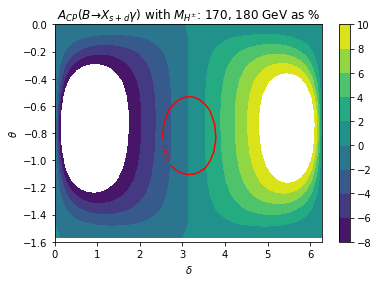}
    }
    \end{subfigure}
    \caption
    { $\mathcal{A}_{CP} ({\overline B} \to X_{s+d} \gamma)$ (as a percentage) in the plane [$\delta, \theta$] with $m_{H^{\pm}_1} = 170$ GeV, $m_{H^{\pm}_2} = 180$ GeV, 
    $\tan\beta =35$ and
    $\tan\gamma =1.32$. Inside the red circle the predicted BR$({\overline B}\to X_s\gamma)$ satisfies 
the current experimental constraint. The white regions are excluded by the current ($3\sigma$) experimental limits on $\mathcal{A}_{CP} ({\overline B} \to X_{s+d} \gamma)$.
    } \label{brcbcs80}
\end{figure}

In Figs.~7a, 7b, and 8
the contours of the CP asymmetries are shown in the plane $[\delta, \theta]$. The other parameters are fixed as $m_{H^{\pm}_1}=170$~GeV, $m_{H^{\pm}_2}=180$~GeV, $\tan\beta=35$, and $\tan\gamma=1.32$. 
The scale $\mu_b$ and the hadronic parameters are taken to be the same as 
in Figs.~5a, 5b and 6.
Inside the red circles the predicted BR$({\overline B}\to X_s\gamma)$ satisfies 
the current ($3\sigma$) experimental constraint, and restricts the allowed parameter space to be roughly
$2.5 < \delta < 3.5$ and $-0.5 < \theta< -1.1$ (i.e. an ellipse centred on around $\delta=3$). The white regions in all plots violate the current $3\sigma$ experimental limits (see Table~II) on the displayed asymmetry. 
In Fig.~7a it can be seen that roughly the right half ($\delta>3$) of the ellipse is ruled out from $\mathcal{A}_{CP} ({\overline B} \to X_s \gamma )$.
In Figs.~7b and 8, in the allowed region of the plane $[\delta, \theta]$ the asymmetries increase in magnitude as $\delta$ is varied from $\delta=\pi$ to $\delta\approx 2.5$, and values of  $\Delta A_{CP}\approx -1.5\%$ and
$\mathcal{A}_{CP} ({\overline B} \to X_{s+d} \gamma)\approx -3\%$ can again be reached.
The theoretical uncertainty is significant, and will be quantified in what follows.


We now consider the theoretical uncertainty of our predictions that arise from varying
the scale $\mu_b$ and the hadronic parameters.
In Tabs.~V, VI and VII the parameters are fixed as $m_{H^{\pm}_1}=170$~GeV, 
$m_{H^+_2}=180$~GeV, 
$\theta=-\frac{\pi}{4}$ and $\delta=2.64$ (same as in Figs.~5a, 5b and 6);
$\tan\beta=35$ and $\tan\gamma=1.32$ (same as in Figs.~7a, 7b and 8). 
Tab.~V uses the lowest possible values of the hadronic parameters, Tab.~VI uses the central values, and Tab. ~VII uses the maximum values.
In each table the value of the scale $\mu_b$ is taken to be  $\mu_b=m_b/2$, $m_b$ and $2m_b$.  The pole $b-$quark mass is $4.77\pm 0.06$ GeV, and
in Tabs.V, VI and VII we take 4.71 GeV, 4.77 GeV and 4.83 GeV respectively.
This scale dependence corresponds to the NNLO contributions 
in BR$({\overline B}\to X_s\gamma)$ and the NLO contributions in the CP asymmetries. 
The uncertainty from $\mu_b$ is around 50~\% for $\Delta A_{CP}$ and $\mathcal{A}_{CP} ({\overline B} \to X_{s+d} \gamma)$ in each table.
One can see that increasing the scale $\mu_b$ makes both $\Delta A_{CP}$ and $\mathcal{A}_{CP} ({\overline B} \to X_{s+d} \gamma)$ more negative.
The CP asymmetry $A_{CP}({\overline B}\to X_s\gamma)$ is very significantly affected by the
change of the hadronic parameters, so that even the sign of the asymmetry is flipped. 
The effect of the change of the hadronic parameters on $\Delta A_{CP}$ is also severe (due to it being proportional to $\tilde{\Lambda}_{78}$), while the effect on
 $\mathcal{A}_{CP} ({\overline B} \to X_{s+d} \gamma)$ is less significant.
 The maximum and minimum values of the observables in Tabs.~V, VI and VII are as follows:
\begin{align}
&2.724\times 10^{-4}  <\text{BR} ({\overline B} \to X_s \gamma) < 2.968\times 10^{-4} \;,
\\
&-4.137  \% < \mathcal{A}_{CP} ({\overline B} \to X_s \gamma )  < 0.581 \%\;,
\\
&-1.785   \% <  \Delta A_{CP}  < -0.111 \%\;,
\\
&-3.323  \% <  \mathcal{A}_{CP} ({\overline B} \to X_{s+d} \gamma) < -0.974   \%\;.
\end{align} 
We note that a full scan over the hadronic parameters might result in larger asymmetries.

\begin{table}
\centering
\begin{tabular}{|c|c|c|c|c|}

	\hline
	 $\mu_b$ & ${\overline B} \to s \gamma $ $(\times 10^{-4})$  &$\mathcal{A}_{CP} (\bar{B} \to X_s \gamma )$\% & $\Delta A_{CP}$ \%&$\mathcal{A}_{CP} ({\overline B} \to X_{s+d} \gamma)$
\%  \\	 

	 \hline
$m_b/2$ & $2.912$ & $-3.170$ & $-0.111$ &  $-0.974$  \\
$m_b$ & $2.968$ & $-3.636$ & $-0.134$ &  $-1.058$  \\
$2m_b$ & $2.801$ & $-4.137$ & $-0.163$ &  $-1.153$  \\
\hline
\end{tabular}
\caption{Dependence of the asymmetries on the scale $\mu_b$, taking the lowest values of the hadronic parameters and $m_b = 4.71 $GeV. Parameters are fixed as follows: $m_{H^{\pm}_1} = 170$ GeV, $m_{H^{\pm}_2} = 180$ GeV,
$\theta = - \frac{\pi}{4}$, $\tan\beta = 35$, $\tan \gamma = 1.32$, $\delta = 2.64$, $\tilde{\Lambda}^{u}_{17} = -0.66$ GeV, $\tilde{\Lambda}^{c}_{17} = -0.007$ GeV, $\tilde{\Lambda}_{78} = 0.017$ GeV with LO $C_7, C_8$. }
\end{table}

\begin{table}
\centering
\begin{tabular}{|c|c|c|c|c|}
	\hline
	 $\mu_b$ & ${\overline B} \to s \gamma $ $(\times 10^{-4})$  &$\mathcal{A}_{CP} (\bar{B} \to X_s \gamma )$\% & $\Delta A_{CP}$ \%&$\mathcal{A}_{CP} ({\overline B} \to X_{s+d} \gamma)$ \%  \\	 
	 \hline
$m_b/2$ & $2.888$ & $-1.220$ & $-0.562$ & $-1.755$ \\
$m_b$ & $2.931$ & $-1.663$ & $-0.673$ & $-2.151$  \\
$2m_b$ & $2.761$ & $-2.212$ & $-0.820$ & $-2.670$ \\
	\hline
\end{tabular}
\caption{Dependence of the asymmetries on the scale $\mu_b$, taking the central values of the hadronic parameters and $m_b = 4.77$ GeV. Parameters are fixed as follows: $m_{H^{\pm}_1} = 170$ GeV, $m_{H^{\pm}_2} = 180$ GeV,
$\theta = - \frac{\pi}{4}$, $\tan\beta = 35$, $\tan \gamma = 1.32$, $\delta = 2.64$, 
$\tilde{\Lambda}^{u}_{17} = 0$ GeV, $\tilde{\Lambda}^{c}_{17} = 0.0085$ GeV, $\tilde{\Lambda}_{78} = 0.0865$ GeV with LO $C_7, C_8$.  }
\end{table}
\begin{table}[h]
\centering
\begin{tabular}{|c|c|c|c|c|}
	\hline
	 $\mu_b$ & ${\overline B} \to s \gamma $ $(\times 10^{-4})$  &$\mathcal{A}_{CP} (\bar{B} \to X_s \gamma )$\% & $\Delta A_{CP}\%$&$\mathcal{A}_{CP} ({\overline B} \to X_{s+d} \gamma)$ \%  \\	 
	 \hline
$m_b/2$ & $2.865$ & $1.145$ & $-1.223$ & $-2.123$ \\
$m_b$ & $2.896$ & $0.914$ & $-1.466$ & $-2.641$ \\
$2m_b$ & $2.724$ & $40.581$ & $-1.7854$ & $-3.323$ \\
	\hline
\end{tabular}
\caption{Dependence of the asymmetries on the scale $\mu_b$, taking the largest values of the hadronic parameters and $m_b = 4.83$ GeV. Parameters are fixed as follows: $m_{H^{\pm}_1} = 170$ GeV, $m_{H^{\pm}_2} = 180$ GeV,
$\theta = - \frac{\pi}{4}$, $\tan\beta = 35$, $\tan \gamma = 1.32$, $\delta = 2.64$
$\tilde{\Lambda}^{u}_{17} = 0.66$ GeV, $\tilde{\Lambda}^{c}_{17} = 0.010$ GeV, $\tilde{\Lambda}_{78} = 0.19$ GeV with LO $C_7, C_8$.   }
\end{table}	

\subsection{Electric dipole moments, collider limits and theoretical consistency}
\noindent
In a separate publication \cite{EDMs}, some of us addressed the calculation of both the neutron and EDMs in the 3HDM discussed here, as these observables will be affected by a non-zero value
of the CP violating (CPV) phase $\delta$.
Without pre-empting the results to appear therein,  it has been checked that the regions of 3HDM parameter space explored in our present analysis are generally compliant with constraints coming from both neutron and electron EDMs. However, some regions of the parameter space covered here would be excluded. Specifically, with reference to the $\tan\beta$ and $\tan\gamma$ values adopted  and the 
$[\delta, \theta]$ plane considered, we can anticipate that the regions centred around $\theta\approx -0.8$ and $\delta \approx 1.4$ and 4.6 would be excluded by the combination of the two EDMs. However, the expanse of such an invalid  parameter space diminshes significantly as $m_{H_1^\pm}$ and $m_{H_2^\pm}$ get closer, to the extent that no limits can be extracted from such observables in the case of exact mass degeneracy of the two charged Higgs states, for suitable values of their Yukawa couplings. Hence, the majority of the results presented here are stable against EDM constraints.
Indeed, it should further be noted that both in the present paper and in Ref. \cite{EDMs}, for computational reasons, the neutral Higgs sector of the 3HDM has essentially been decoupled. Hence, in the case of a lighter
neutral scalar spectrum 
one may potentially invoke cancellations occurring between the  charged and neutral Higgs boson states (including the SM-like one) of the CPV 3HDM (in the same spirit as those of Ref.~\cite{Kanemura:2020ibp} for the CPV Aligned 2HDM), which could further reduce the impact of EDM constraints.
Moreover, one also ought to make sure that the $H_1^\pm$ and $H_2^\pm$ spectra of masses and couplings adopted here do not violate bounds
coming from colliders, specifically LEP/SLC, Tevatron and the LHC. Again, based on the forthcoming results of Ref.~\cite{EDMs}, we anticipate this being the case in the present context. 
Finally, in Ref.~\cite{EDMs}, it will also be shown that the values of the Yukawa parameters adopted in this paper are compliant with theoretical self-consistency requirements of the 3HDM stemming from vacuum stability and perturbativity.

\section{Conclusions}  
\noindent
In the context of 3HDMs with NFC the magnitudes of three CP asymmetries that involve the decay $b\to s/d\gamma$ have been studied.
In the SM, the CP asymmetry in the inclusive decay ${\overline B}\to X_s\gamma$ alone (${\mathcal{A}}_{X_s \gamma}$) has a theoretical error from long-distance contributions that render it unlikely to
provide a clear signal of physics beyond the SM at the ongoing BELLE II experiment.
The untagged asymmetry ($\mathcal{A}_{CP} ({\overline B} \to X_{s+d} \gamma)$) and the difference of CP asymmetries ($\Delta\mathcal{A}_{X_{s}\gamma}$) are both predicted to be essentially zero in the SM, with
negligible theoretical error. Hence these latter two observables offer better prospects of revealing new physics contributions to $b\to s/d\gamma$.

In the context of 3HDMs there are two charged scalars that contribute to the process $b\to s/d\gamma$. 
There are six new physics parameters (two masses of the charged scalars, and four parameters that determine the Yukawa couplings of the charged scalars) that together 
enable the relevant Wilson coefficients to contain a sizeable imaginary part. 
In three of the five types of 3HDM the magnitude of $\mathcal{A}_{CP} ({\overline B} \to X_{s+d} \gamma)$ and $\Delta\mathcal{A}_{X_{s}\gamma}$
can reach values such that a $5\sigma$ signal at the BELLE II experiment with 50 ab${^{-1}}$  of integrated luminosity would be possible.
Although the parameter space for such a clear signal is rather small (which is also usually the case in other models of physics beyond the SM), it was shown that a 3HDM could accommodate any such 
signal, and thus would be a candidate model for a statistically significant excess (beyond the SM prediction) in these asymmetry observables.

\section*{Acknowledgements}
\noindent
AA and SM are supported in part through the STFC Consolidated Grant ST/L000296/1. 
SM is supported in part through the NExT Institute. SM and MS acknowledge the H2020-MSCA-RISE-2014 grant no. 645722 (NonMinimalHiggs). 
TS is supported in part by JSPS KAKENHI Grant Number 20H00160. 
TS and SM are partially supported by the Kogakuin University Grant for the project research 
"Phenomenological study of new physics models with extended Higgs sector”. We thank H. E. Logan and D. Rojas-Ciofalo for reading the manuscript and for useful comments.


\end{document}